\documentclass{svproc}
\usepackage{url}

\usepackage{amsmath}
\usepackage{graphicx}
\usepackage{float}

\usepackage[T1]{fontenc}

\usepackage{lmodern}
\usepackage{tgcursor}

\usepackage{listings}
\begin{document}
\mainmatter              
\title{Bayes Linear Emulation of Simulated Crop Yield}
\titlerunning{Bayes Linear Emulation}  
%
\author{Muhammad Mahmudul Hasan \and Jonathan A. Cumming}
\authorrunning{Muhammad M Hasan et al.} 
%
\tocauthor{Muhammad Mahmudul Hasan, Jonathan A. Cumming}
\institute{Department of Mathematical Sciences, Durham University, Durham,  UK\\
\email{muhammad.m.hasan@durham.ac.uk}\\
\email{j.a.cumming@durham.ac.uk}
}

\maketitle              

\begin{abstract}
The analysis of the output from a large-scale computer simulation experiment can pose a challenging problem in terms of size and computation. We consider output in the form of simulated crop yields from the Environmental Policy Integrated Climate (EPIC) model, which requires a large number of inputs - such as fertiliser levels, weather conditions, and crop rotations - inducing a high dimensional input space. In this paper, we adopt a Bayes linear approach to efficiently emulate crop yield as a function of the simulator fertiliser inputs. We explore emulator diagnostics and present the results from emulation of a subset of the simulated EPIC data output.
\end{abstract}
\section{Introduction}
The use of computer simulations to model complex systems has become increasingly popular, with applications spanning many areas of scientific study. The Environmental Policy Integrated Climate (EPIC) model \cite{will:epic} is one such example, constructed to explore and simulate the behaviour of various crops over time in response to key inputs such as crop rotation, fertilizer levels, land management, weather conditions, and other environmental variables. Consideration of the simulator behaviour in response to such a large number of inputs is a challenging and high-dimensional problem. An effective strategy to model such simulator output is through the use of an emulator, which acts as a statistical surrogate for the computationally expensive computer simulation. In this paper, we explore emulator construction and diagnostics for a subset of output from the EPIC simulation, with a view to developing our understanding of yield trend corresponding to fertilizers inputs.

\section{The Simulator}
The scope and scale of EPIC's behaviour and outputs are quite complex, encompassing: (i) crop rotation - the sequence of crops planted and harvested during the run of the simulation; (ii) fertilizer inputs - the levels of nitrogen and phosphorous applied during the simulation; (iii) land characteristics - encapsulated in the form of different soils and land steepness; and (iv) weather - while not an input that can be controlled, a variety of historical weather scenarios are applied. Evaluation of the simulator then outputs a time series of various attributes relating to the crop and land conditions throughout a 60-year simulation. For this analysis, we focus on the annual reported yield for each of the crops of interest, thus our simulated data reduces to the form of a single simulated yield for each combination of input variables.
For our analysis, we will focus on the continuous fertiliser inputs of nitrogen ($N$) and phosphorus ($P$) levels, which were simulated over a discrete grid of values, each with 13 values over [0, 100]. Thus, for a given crop and fixed combination of land and weather variables, we obtain a grid of 169 simulated yields, $Y$, in response to $N$ and $P$. Examples of such yield responses are shown in Fig. 1, plotting only the marginal response of spring barley yield to $N$. We note the common feature of a monotone increasing yield in response to increased fertilisation, however, we observe a number of simulations deviate from this expected pattern being either constant or changing suddenly.
 \vspace{-20pt}
 \begin{figure}
\centering
\includegraphics[width=12cm,height=4cm]{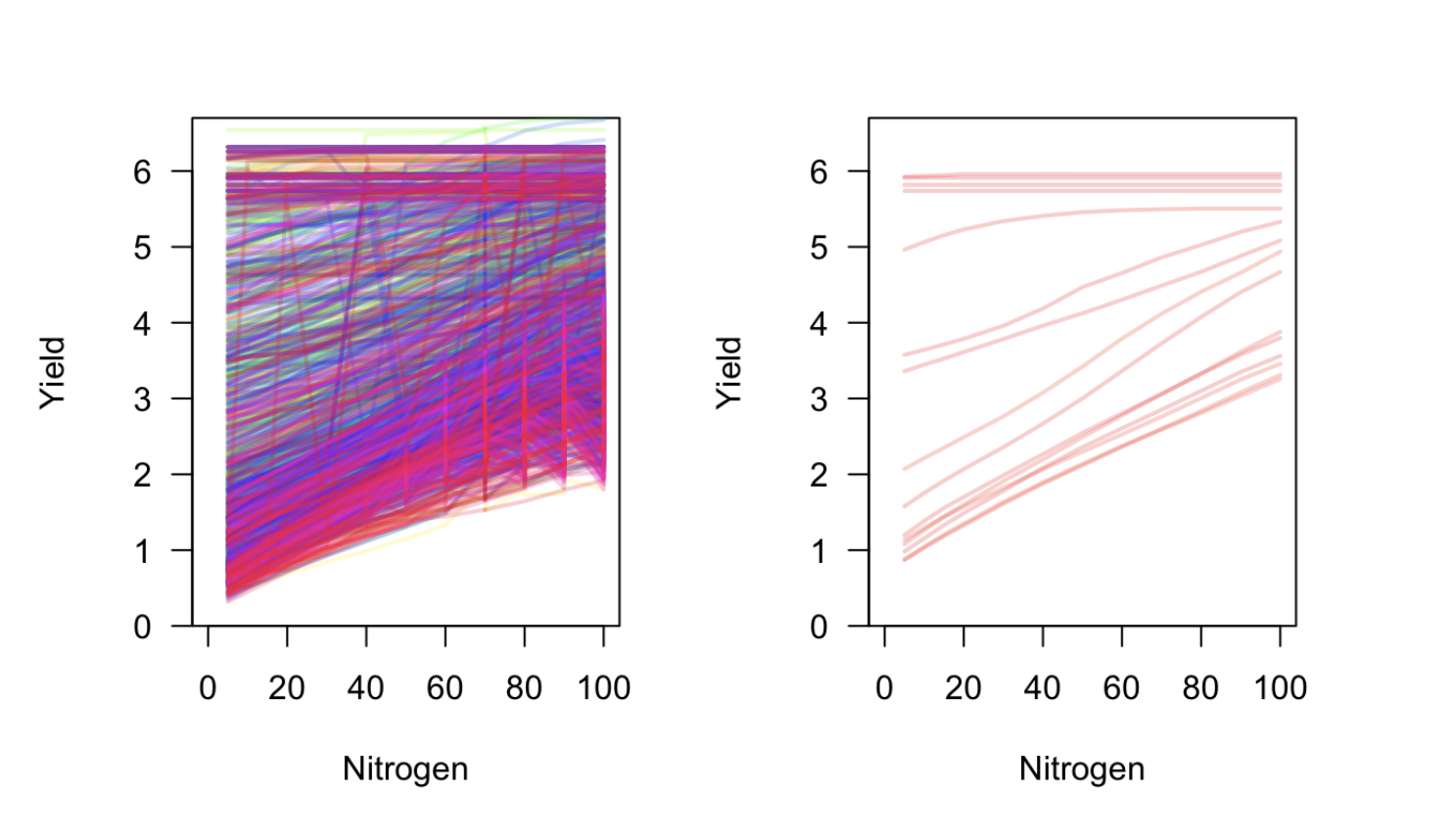}
\caption{Left panel: Line plot for the 1250 unique combinations of spring barley data with respect to nitrogen fertilizer; Right panel: Line plot for 15 different small unique combinations to assess the clear trend of the data.}
\end{figure}
 \vspace{-20pt}
\section{Methodology}
\subsection{Bayes linear methods}
In Bayesian inference, data is used to update beliefs about random quantities of interest - typically represented via probability distributions. The Bayesian approach typically combines a parametric model for the data in the form of the likelihood with a prior probability distribution over the model's parameters to produce a posterior distribution for those parameters given the data. This fully probabilistic approach poses a number of practical challenges: the first being making a meaningful prior specification, and the second being computational issues in obtaining or simulating from the posterior distributions.

An alternative approach follows the concept of de Finetti \cite{Finetti:bd}, where we consider belief specifications in terms of expectation rather than probability. This is known as a \emph{Bayes linear approach}, and it operates based on means and variances. The fundamental Bayesian update can then be compactly expressed by the following two equations \cite{Goldstein:wood}:
\begin{eqnarray}
    E_D [B]&=&E[B]+Cov[B,D] Var[D]^{-1} (D-E[D])\label{1}\\
    Var_D [B]&=&Var[B]-Cov[B,D] Var[D]^{-1}  Cov[D,B]\label{2}
\end{eqnarray}
where prior beliefs about random quantities $B$ are updated given data $D$. A key advantage to this approach to Bayesian inference is the elimination of the need for complex sampling schemes to investigate the posterior distributions, as posterior means and variances can be found directly via the equations above. Additionally, without the need to specify full distribution forms for our priors the task of the prior specification is also less complex. However, the key consequence of this and the primary limitation of this approach is that we have potentially sacrificed the richness of the information provided by operating with probability distributions.

\subsection{Emulation}
For our problem, we will treat each collection of simulated yields as functions of $N$ and $P$. As the functional form of the simulated yield response to $N$ in Fig. 1 is not consistent with a single parametric model, we will use a statistical emulator to model the relationship between the inputs from the simulator to the outputs. The general form of an emulator is based around a combination of a regression surface with correlated and uncorrelated errors as follows:
\begin{align*}
f(x)=\sum_{i=1}^{p}\beta_ig_i(x)+ u(x)\tag{3}
\end{align*}
where $\sum_{i=1}^{p}\beta_ig_i(x)$ represents the mean function in a regression form, expressed in terms of the input variables, $x$. The parameters $\beta_i$ are unknown scalar regression coefficients corresponding to the regression basis functions for the active inputs $g_i (x)$. The final component is then $u(x)$, which is a zero-mean weakly stationary process to explain additional variation around the mean function in terms of $x$. To complete the model specification, we require a covariance function for the residual process $u(x)$, which typically has a Gaussian form:
 \begin{align*}
Cov[u(x_1 ),u(x_2 )]= \sigma^2  \exp[-\theta|x_1-x_2|^2]\tag{4}
\end{align*}
for any pair of inputs $x_1$ and $x_2$, with a correlation length parameter $\theta$ and variance $\sigma^2$. To construct our Bayes linear emulator \cite{Cumming:Goldstein}, we structure the mean function as a simple regression in terms of the simple basis $[1,N,P]$. Thus, from the equation (3) we can write as $E[f(x)]=E[\beta_0 ]+E[\beta_1 ]N+E[\beta_2 ]P$, in terms of three regression coefficients $\beta_0,\beta_1,\beta_2$. Prior expectations and variances for these coefficients and $\sigma^2$ were assigned to the corresponding least-squares estimates over the training data set.  The correlation length parameters for $N$ and $P$ were assigned to a value of 0.015 after investigation via cross-validation, and to reflect a common level of smoothness of yield in response to both inputs.

\subsection{Emulator diagnostics}
To assess the quality of our emulator, we can compute various diagnostics. First, the resolution \cite{Cumming:Goldstein} of the Bayes linear update can be expressed as,
\begin{align*}
     R_D[f(x)]=1-\frac{Var_D [f(x)]}{Var[f(x)]}\tag{5}
\end{align*}
The resolution lies between 0 and 1, and functions much like a classical  $R^2$ where that resolution values close to 1 indicate a high proportion of the variation has been explained. Secondly, the standardized prediction errors (SPE) \cite{Cumming:Goldstein} for a simulation value $Y$ with corresponding inputs $x$ can be expressed as,	
\begin{align*}
    SPE=\frac{Y-E_D [f(x)]}{\sqrt{Var_D (f(x))}}\tag{6}
\end{align*}
Large values of SPE indicate a clear conflict between the emulator and simulator, indicative of deficiencies in the fit of the emulator or surprising simulator output values. In general, values of (6) of absolute value greater than $3$ are used to identify such problems \cite{Cumming:Goldstein}.
\vspace{-5pt}

\section{Results and Discussion}
For our analysis, we present results from a subset of crops, namely maize and spring barley. For each crop, our simulations are structured we have 1250 simulations of yield corresponding to different simulation conditions, each of which explores a 13Ã-13 grid of combinations of the two fertiliser inputs, $N$ and $P$. Focussing on a single grid of simulated yield, we construct a Bayes linear emulator based on a simple regression (3) and a correlated error with covariance function (4) using 80\% of the available data, reserving the remaining 20\% for testing and diagnostics.\

The emulator is updated from the training data via the Bayes linear formulae (1) and (2), with results shown in Fig. 2. The left and centre panels show the emulated mean maize yield and its associated standard deviation as functions of $N$ and $P$.  We note that the crop yield is increasing with increasing Nitrogen levels, though the effect of Phosphorous is much less pronounced and arguably only important when levels of Nitrogen are low. The standard deviation plot highlights low levels of uncertainty in mean maize yield around the locations for which we have simulations, with uncertainty increasing as we move away from these points. The right panel shows the emulated mean yield for spring barley, where we note that the weak dependency on Phosphorous has now disappeared entirely and the crop yield appears insensitive to values of $P$.\

Diagnostic plots for the emulation of maize yield are given in Figure 3. The plot of the emulator resolution (left) displays high values (greater than 0.7) over much of the space, indicating the emulator has explained much of the data variability. The blue regions of low resolution indicate locations corresponding to the test data, which were not used for emulator fitting, hence little data was available to reduce the variance in these locations. Standardised prediction errors are shown in the centre and right panels of Fig. 3, and we note that all points lie within Â±2, suggesting a high degree of consistency and agreement between emulator and simulator.

 \vspace{-22pt}
\begin{figure}
    \centering
    \includegraphics[width=12cm,height=4cm]{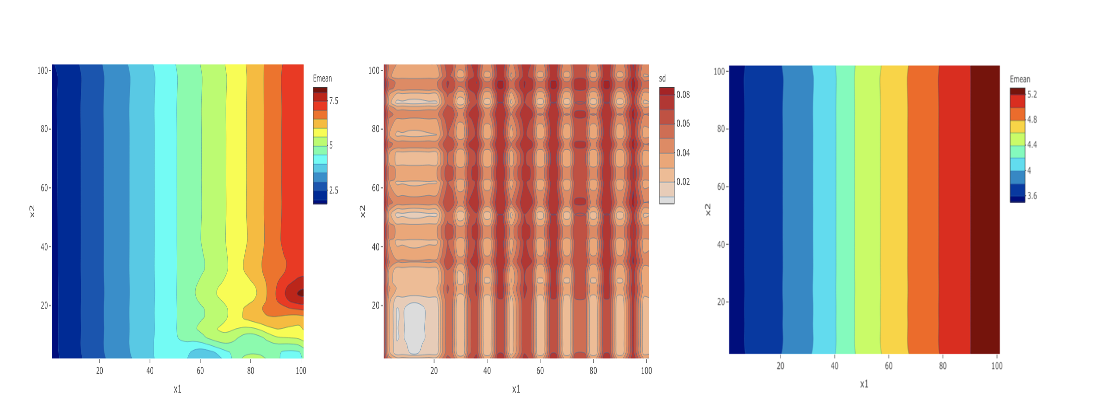}
    \caption{Left: emulated mean maize yield as a function of Nitrogen ($x_1$) and Phosphorus ($x_2$); middle: emulator standard deviation for maize yield; right: emulated mean spring barley yield.}
\end{figure}
\vspace{-30pt}
\begin{figure}[H]
     \centering
    \includegraphics[width=12cm,height=4cm]{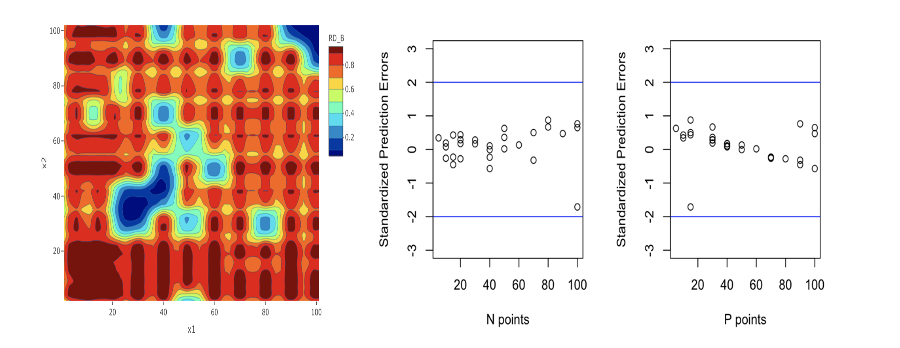}
   \caption{From left to right: plot of resolution (5) of the Bayes linear update; standardised prediction error for test data on the maize yield emulator as a function of $N$; standardised prediction error for test data on the maize yield emulator as a function of $P$.}
\end{figure}

\section{Concluding Remarks}
Emulation is an effective tool for modelling computer simulations, where a parametric model of the simulator's response to changes in inputs may not be known a priori. For data such as these EPIC simulations, the shape of the yield response was not consistent between simulations necessitating an emulation-based approach where the yield response could be determined from prior information and the simulation output itself. A fully Bayesian approach would require distributional specifications for each of the parameters in such an emulator and simulation-based methods for any subsequent inference, which becomes challenging when dealing with computer models with large numbers of outputs. For our analysis, we adopted a Bayes linear approach which simplified the computation and complexity in fitting the model substantially, while still providing a powerful tool for modelling and analysing the computer model output. Additionally, the emulator's quality and performance can be readily assessed and monitored through the use of appropriate diagnostics. Looking ahead, a natural progression from this work is to broaden the input space and consider the effects of the entire collection of simulator inputs - including both continuous and categorical variables.

%
%

 \section{Appendix }
 
\subsection{R code for Bayes Linear Emulation}

R code implementation of our methods is available at \url{https://sites.google.com/d/1siopGfjK_btE99h2TU2BGfwzgEn8szeA/p/1NjQxo2ti4d23Hc8zIeXP0Py0NsXVzHpK/edit}.

\end{document}